\documentclass[12pt]{article}

\newcommand{\ra}{\rightarrow}

\newcommand{\vareps}{\varepsilon}

\newcommand{\norm}[1]{\|#1\|}

\newcommand{\Ima}{{\rm Im}}
\newcommand{\Rea}{{\rm Re}}
\title{\vspace{-1in}\parbox{\linewidth}{\small\hfill\shortstack{IASSNS-HEP-97/124\\
CALT-68-2143}}
\vspace{0.6in}\\
Singular Monopoles and Supersymmetric Gauge Theories in Three Dimensions}
\author{Sergey A.  Cherkis\thanks{Research supported in part by DOE grant 
DE-FG03-92-ER40701}\\
{\it\small California Institute of Technology}\\
{\it\small  Pasadena, CA 91125}
\and Anton Kapustin\thanks{Research supported in part by DOE grant DE-FG02-90-ER40542}\\
{\it\small School of Natural Sciences, Institute for Advanced Study}\\
{\it\small Olden Lane, Princeton, NJ 08540}}

\begin{document}
\begin{titlepage}
\renewcommand{\thepage}{ }
\renewcommand{\today}{ }

\thispagestyle{empty}
\maketitle

\begin{sloppypar}
\begin{abstract} According to the proposal of Hanany and Witten, Coulomb
branches of $N=4$ $SU(n)$ gauge theories in three dimensions are isometric to 
moduli spaces of BPS monopoles.  We generalize this proposal to gauge theories
with matter, which allows us to describe the metrics on their spaces of vacua by
means of the hyperk\"ahler quotient construction.  To check the identification
of moduli spaces a comparison is made with field theory predictions.  For
$SU(2)$ theory with $k$ fundamental hypermultiplets the Coulomb branch is
expected to be the $D_k$ ALF gravitational instanton, so our results lead to a
construction of such spaces.  In the special case of $SU(2)$ theory with four or fewer
fundamental hypermultiplets we calculate the complex structures on the moduli
spaces and compare them with field-theoretical results.
We also discuss some puzzles with brane realizations of three-dimensional
$N=4$ theories.

\end{abstract}
\end{sloppypar}
\end{titlepage}

\section{Introduction} 

Realizing supersymmetric gauge theories as theories on
D-branes proved to be very useful for identifying their excitations and
spaces of vacua.  In certain cases this approach allows one to show that the
Coulomb branch of the space of vacua is the same as the moduli space of some
self-dual Yang-Mills configurations in an auxiliary gauge theory.  For example,
as described in Ref.~\cite{HW}, the Coulomb branch of $N=4$ supersymmetric $SU(n)$
Yang-Mills theory in three dimensions is the (centered) moduli space of $n$ $SU(2)$
monopoles.  There are powerful mathematical methods, such as twistor methods and
the ADHM-Nahm construction, developed to describe solutions of the self-duality
equation.  Using these methods one can compute the metric on the space of vacua.
In addition, realization of the same theory by different D-brane configurations
clarifies the connection between different mathematical constructions and yields
nontrivial predictions about the geometry of the space of vacua.

In this paper we generalize the correspondence between the Coulomb branches and
monopole moduli spaces to the case of $D=3, N=4$ $SU(n)$ gauge theory with $k$
hypermultiplets in the fundamental representation.  We show that the Coulomb
branch is given by the (centered) moduli space of $U(2)$ ``monopoles'' on the $A_{k-1}$
ALF space (also called multi-Taub-NUT), as described in Section~\ref{alfmon}.
These ``monopoles'' turn out to be equivalent to solutions of $U(2)$ Bogomolny equations
on ${\bf R}^3$ with $k$ singularities (Section~\ref{monsing}). They can be described by the
spectral data in the twistor description~\cite{KrTh,us} or by solutions of
Nahm equations.

In the strong coupling limit (i.e. $g\ra\infty$) the metric on the Coulomb 
branch of the $SU(n)$ gauge theory with $k$ hypermultiplets is identical to
the metric on the Higgs branch of the mirror gauge theory described
in Ref.~\cite{Mirror}. The latter does not receive any quantum corrections
and can be easily computed. However, going from $g=\infty$ to finite $g$
is highly nontrivial in general. For example, the Coulomb branch of the
$SU(2)$ gauge theory with $k$ fundamentals is expected to be the ALF 
gravitational instanton of type $D_k$ \cite{S,SW}, while the $g\ra\infty$ limit is
the ALE gravitational instanton. Self-dual ALE metrics have been completely
classified by Kronheimer~\cite{KrALE}, but very little is known about
the ALF case (see though Ref.~\cite{Dancer2}). The monopole methods
allow us to construct $D_k$ ALF metrics as (infinite-dimensional) hyperk\"ahler
quotients.

The brane configurations which we use are the configurations 
described in Ref.~\cite{HW}; they contain both D and NS5-branes.
If one moves D-branes around, naively there seem to be phase transitions in the
worldvolume gauge theory, with new matter multiplets appearing when 
D5 and NS5-branes cross. In Ref.~\cite{HW} it was argued that phase transitions are
actually absent.  In some cases this can be explained by creation of a
D3-brane when a D5 and an NS5 cross.  However there are situations when brane
creation does not help and even complicates the picture.  In Section~\ref{phase}
we analyze such situations in terms of monopole moduli spaces and argue that
phase transitions are still absent.  For this to work, certain states of
fundamental strings stretched between D-branes must be absent, although simply
drawing the picture suggests otherwise.

In Section~\ref{moduli} we construct the moduli spaces of singular monopoles by
means of Nahm equations. We work out in detail the examples corresponding to $SU(2)$
gauge theory with up to four fundamental hypermultiplets in Section~\ref{example}.
In Section~\ref{complex}
we present the complex structures on these spaces and compare them with the
Seiberg-Witten solutions of the corresponding four-dimensional theories. 

In the Appendix we discuss the properties of solutions of self-duality equations
on ${\bf R}^3\times {\bf S}^1$ and on $A_{k-1}$ ALF space needed for our discussion
of phase transitions in Section~\ref{phase}.


\section{Brane Configurations} \label{alfmon} 

Following Ref.~\cite{HW} we
consider configurations of D3, D5 and NS5-branes in IIB string theory which leave
eight supersymmetries unbroken. Let two parallel NS5-branes be some distance
$d$ apart in the $x^6$ direction with worldvolumes parallel to $x^0, x^1, x^2, x^3,
x^4, x^5$. Let $n$ D3-branes stretch between them in the $x^6$ direction, with
other directions of D3's being parallel to $x^0,x^1,x^2$.  This configuration of
branes is illustrated in Figure~1(a).  The theory on D3-branes reduces to the
$D=3,\ N=4$\ $U(n)$ Yang-Mills theory in the infrared limit.  Every such
configuration of branes corresponds to a particular vacuum of the Yang-Mills
theory.  As described in Ref.~\cite{HW}, D3-branes look like $SU(2)$ monopoles
in the
$x^3, x^4, x^5$ directions in the theory on the NS5-branes.  Vacua of
the $U(n)$ Yang-Mills theory on D3-branes are in one-to-one correspondence with
charge $n$ monopoles on the NS5-branes. In order to describe the $SU(n)$ Yang-Mills
theory we should fix the center of mass of the D3-branes. Thus vacua of this theory
are given by ``centered'' monopoles.

\begin{figure}

\setlength{\unitlength}{0.9em}
\begin{center}
\begin{picture}(34,13)
\put(8.5,0.5){\makebox(0,0){(a)}}
\put(6,2){\line(0,1){10}}
\put(11,2){\line(0,1){10}}
\multiput(6,5)(0,0.5){3}{\line(1,0){5}}
\put(6,9){\line(1,0){5}}
\multiput(8.5,7)(0,0.5){3}{\circle*{.1}}

\put(25.5,0.5){\makebox(0,0){(b)}}
\put(23,2){\line(0,1){10}}
\put(28,2){\line(0,1){10}}
\multiput(23,5)(0,0.5){3}{\line(1,0){5}}
\put(23,9){\line(1,0){5}}
\multiput(25.5,7)(0,0.5){3}{\circle*{.1}}

\put(28,9.5){\line(1,0){4}}\put(32,9.5){\makebox(0,0){\boldmath $\times$}}
\put(28,8){\line(1,0){2}}\put(30,8){\makebox(0,0){\boldmath $\times$}}
\put(28,4){\line(1,0){3}}\put(31,4){\makebox(0,0){\boldmath $\times$}}
\put(23,10){\line(-1,0){4}}\put(19,10){\makebox(0,0){\boldmath $\times$}}
\put(23,4){\line(-1,0){4}}\put(19,4){\makebox(0,0){\boldmath $\times$}}
\put(23,3){\line(-1,0){3}}\put(20,3){\makebox(0,0){\boldmath $\times$}}
\multiput(29,5.5)(0,0.5){3}{\circle*{.1}}
\multiput(21,6.5)(0,0.5){3}{\circle*{.1}}

\end{picture}\end{center}
\caption{(a) Parallel NS5-branes (vertical lines) with D3-branes
(horizontal lines) suspended between them. The horizontal
direction corresponds to $x^6$, while the vertical direction 
corresponds to $x^3,\,x^4,$ and $x^5$ collectively. (b) The same,
with additional D5-branes (crosses) connected by D3-branes to NS5-branes.}

\end{figure}
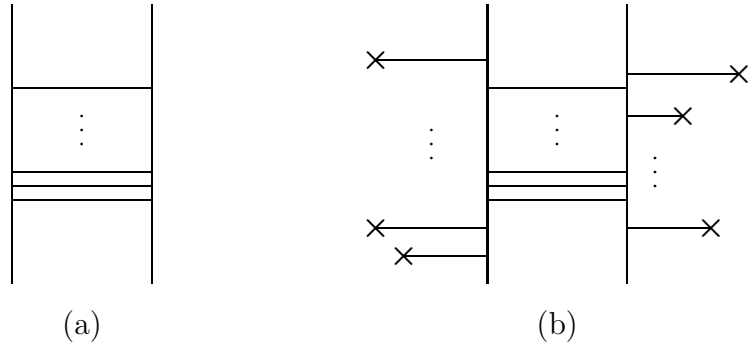

Now let us add $k$ D5-branes stretching along $x^0, x^1, x^2, x^7, x^8, x^9$ and
positioned outside the NS5-branes at points $\vec{p}_\alpha$ in the $(x^3,
x^4, x^5)$ plane (see Figure~1(b)).  Let each D5-brane be connected by one
D3-brane to an NS5-brane closest to it.  We will call these D3-branes external, to
distinguish them from those connecting the two NS5-branes, which we will call
internal.  From the point of view of the internal D3-branes the low-energy
theory is a $U(n)$ gauge theory with $k$ matter hypermultiplets in the
fundamental representation.  Matter comes from the fundamental strings
connecting the internal and external D3-branes.  The question is what this
configuration looks like in the $SU(2)$ theory on the NS5-branes.

To answer the question we perform S and T duality transformations.  First we go
to the S dual picture thus exchanging D5 and NS5-branes.  Then we T dualize
along the $x^6$ direction (after making it periodic) thus turning IIB string
theory into IIA string theory, D5-branes into
D6-branes, and NS5-branes into an $A_{k-1}$ type ALF space.  Tracing the
dualities we have the $A_{k-1}$ ALF metric in $(x^3, x^4, x^5, \hat{x}^6)$, with
$\hat{x}^6$ being the compact direction. Four of the directions of the
D6-branes are wrapped on this space.

What do the D3-branes turn into after the dualities?  If any of the D3-branes
were wrapped around $x^6$, it would turn into a D2-brane located at a
point on the $A_{k-1}$ ALF space. As explained in Ref.~\cite{D}, this D2-brane would
look like an instanton in the $U(2)$ theory on the D6-branes.  To be more
precise, it would be a self-dual $U(2)$ gauge connection on the $A_{k-1}$ ALF space,
somewhat resembling Nahm's calorons~\cite{N}.  Note that the $U(2)$ gauge group
is broken down to $U(1)\times U(1)$ by a nontrivial Wilson line at infinity (in
the original picture this corresponds to a nonzero distance between the 
NS5-branes.) Therefore there may be states in the theory carrying nonzero magnetic
charge. The instanton does not have magnetic charge, and neither does the 
D3-brane wrapped around the $x^6$ direction in the T-dual description.  On the
other
hand, the internal D3-brane does have a magnetic charge, and therefore
corresponds to the monopole solution, by which we mean a self-dual connection
carrying magnetic charge.

If there were no D5-branes in the original brane configuration, we would be
dealing with self-dual connections on ${\bf R}^3\times {\bf S}^1$ rather than on
the $A_{k-1}$ ALF space. Then the internal D3-branes would correspond to
t'Hooft-Polyakov monopoles, i.e. they would not depend on the circle
coordinate~\cite{LeeYi}.  It is highly plausible that this remains true when D5
branes are present.  Indeed, the well-known maxim ``Winding is momentum''
implies in this case that in the IIA picture nothing depends on the $\hat{x}^6$
direction, since nothing is wound around the T-dual direction in the IIB picture.  In
the next section we confirm this by exhibiting monopole solutions on the
$A_{k-1}$ ALF space which do not depend on the circle coordinate.

We still have not discussed the fortune of the external D3-branes after
T-duality.  We postpone the answer until the end of the next section.

\section{Monopoles on $A_{k-1}$ ALF Space} \label{monsing} 

Let us describe precisely what we mean by a monopole on the $A_{k-1}$
ALF space.  In coordinates $x^1, x^2, x^3, \theta$ \ the $A_{k-1}$ ALF
metric is 

\begin{equation} ds^2=V dx^i dx^j\delta_{ij}
+V^{-1}\left(d\theta+\omega^i dx^i\right)^2 ,
\end{equation} 
where $\theta$ has period $4\pi$ and is T-dual to $x^6$, and

\begin{equation}\label{V} V=1+\sum_{\alpha=1}^k
\frac{1}{|\vec{x}-\vec{p}_{\alpha}|}, \ \ grad\ V=curl\ \vec{\omega}.
\end{equation} 
Note that $\omega$ is not a globally defined 1-form; rather it is
a connection on a nontrivial $U(1)$ bundle and can only be defined patchwise.

A $U(2)$ monopole on this space is a smooth self-dual connection $\hat{\bf A}=\hat{A}_0 d\theta+\hat{A}_j dx^j$ on a $U(2)$ vector bundle with a
nontrivial holonomy (Wilson line) at infinity and nonzero magnetic charge,
whose field strength is independent of $\theta$ in some local gauge.  That is,
away from the centers $\vec{x}=\vec{p}_\alpha$ there is a local gauge
transformation $g(\vec{x}, \theta)$, such that $\tilde{A}_0=g^{-1}\hat{A}_0
g-ig^{-1}\partial_{\theta} g$ and $\tilde{A}_j=g^{-1}\hat{A}_j g-ig^{-1}\partial_j g$ are 
independent of $\theta.$

As $\hat{\bf A}$ is smooth and the norm of $\frac{\partial}{\partial\theta}$
vanishes when $\vec{x}\ra\vec{p}_\alpha$, one necessarily has
$\hat{A}_0(\vec{p}_{\alpha})=0$.  $g(\vec{x}, \theta)$ approaches a circle
action with integer weights $\ell_{\alpha}, \ell'_{\alpha}$ near the centers
$\vec{p}_{\alpha}$.  Then we easily see that after the above-mentioned gauge
transformation the eigenvalues of $\tilde{A_0}$ approach
$\ell_{\alpha}/2,\ell'_{\alpha}/2$ as $\vec{x}\ra \vec{p}_{\alpha}$.

Since in the new gauge ${\bf A}$ does not
depend on $\theta$, one may define new fields on ${\bf R}^3$

\begin{equation}\label{change}
\Phi=V \tilde{A}_0, \ \ \ A_i=\tilde{A}_i-\omega_i\tilde{A}_0.
\end{equation} 
Kronheimer noticed~\cite{KrTh} that these fields satisfy the Bogomolny equation if and only if the
initial ${\bf A}$ is a self-dual connection on the ALF space.  Here $\Phi$ is the Higgs field and
$A_i$ is the gauge potential on ${\bf R}^3$.  From Eq.~(\ref{V}) it is easy
to see that $\Phi$ has singularities at $\vec{x}=\vec{p}_{\alpha}$.  Thus
monopoles on $A_{k-1}$ ALF space are in one-to-one correspondence with monopoles
on ${\bf R}^3$ with singular Higgs field 

\begin{equation}\label{sing}
\Phi\ra \frac{1}{2|\vec{x}-\vec{p}_{\alpha}|}\ {\rm diag}\left(\ell_\alpha,\ell'_\alpha\right) 
\end{equation}
near $\vec{x}=\vec{p}_{\alpha}$ in some gauge.  The asymptotic behavior at infinity is the
same as for ordinary monopoles,
\begin{equation}\label{atinf}
\Phi\ra {\rm diag}\left(1-\frac{n-\sum \ell_\alpha}{2r},-1+\frac{n+\sum\ell'_\alpha}{2r}\right).
\end{equation}
We will call such solutions {\it singular U(2) monopoles}.

What is the meaning of $n$ in Eq.~(\ref{atinf})?
The rank 2 $U(2)$ bundle decomposes
into the sum of eigenspaces of $\Phi$, $E=M\oplus M',$ where $M$ and $M'$ are
line bundles. It follows from Eq.~(\ref{sing}) and 
Bogomolny equations that upon restriction to a small sphere
around $\vec{x}=\vec{p}_\alpha$ the degrees of $M$ and $M'$ are $-\ell_\alpha$ and $
-\ell'_\alpha.$ In other
words, there is a point-like Dirac monopole with charges $-\ell_\alpha,-\ell'_\alpha$ at
$\vec{x}=\vec{p}_\alpha$ embedded in the diagonal subgroup of $U(2)$.
Similarly, the total magnetic charges of the
configuration (the degrees of $M$ and $M'$ restricted to a very large sphere) are
$n-\sum \ell_\alpha$ and $-n-\sum \ell_\alpha$. 
Let us now focus on the $SU(2)$ subgroup of $U(2)$. Then the total magnetic charge in the
$SU(2)$ is $n-\sum(\ell_\alpha-\ell'_\alpha)/2$, while the charge carried by the $\alpha$th
Dirac monopole is $-(\ell_\alpha-\ell'_\alpha)/2.$ 
Therefore $n$ is naturally interpreted as the number of smooth nonabelian monopoles
in the configuration. (Kronheimer~\cite{KrTh} calls it nonabelian charge.) One expects
that $n\geq 0$, and it can be shown~\cite{KrTh} that this is indeed the case.

The Dirac monopole with $\ell=1,\ell'=0$ is in fact the
reincarnation of the right external D3-brane in the initial brane configuration. This becomes
quite obvious if one recalls that the NS5-branes in Figure~1(b) correspond to the
two $U(1)$ factors in the diagonal subgroup of $U(2)$, and that the end of the D3 brane ending
on the NS5-brane from the left (right) carries magnetic charge +1 (-1) in the
corresponding $U(1)$. Similarly, the left external D3-brane maps under $T$-duality to
a Dirac monopole with $\ell=0,\ell'=-1$.

To summarize, the configuration of $n$ internal and $k$ external D3-branes
corresponds to a solution of $U(2)$ Bogomolny equations
with nonabelian charge $n$ and with Higgs field having $k$ singularities
as in Eq.~(\ref{sing}). Depending on whether the $\alpha$th D3-brane is right or left
we have $\ell_{\alpha}=1,\ell'_\alpha=0$ or $\ell_{\alpha}=0,\ell'_\alpha=-1$.

\section{Phase Transitions?}  \label{phase} 

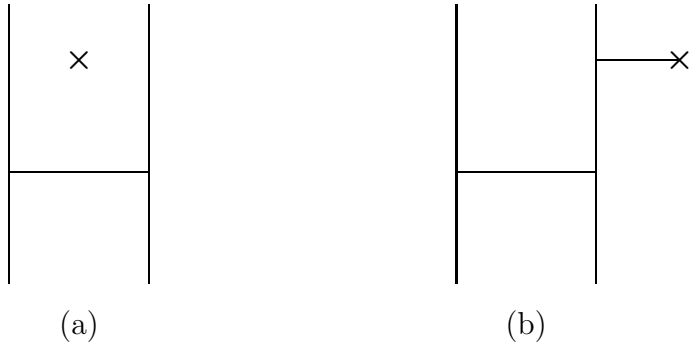
\begin{figure}
\label{pic}
\setlength{\unitlength}{0.9em}

\begin{picture}(34,13)
\put(7.5,0.5){\makebox(0,0){(a)}}
\put(5,2){\line(0,1){10}}
\put(10,2){\line(0,1){10}}
\put(5,6){\line(1,0){5}}
\put(7.5,10){\makebox(0,0){\boldmath $\times$}}

\put(23.5,0.5){\makebox(0,0){(b)}}
\put(21,2){\line(0,1){10}}
\put(26,2){\line(0,1){10}}
\put(21,6){\line(1,0){5}}

\put(26,10){\line(1,0){3}}
\put(29,10){\makebox(0,0){\boldmath $\times$}}

\end{picture}
\caption{Starting with configuration (a) and moving the D5-brane to
the right one gets configuration (b). In both configurations the 
low-energy gauge theory on the D3-brane has one hypermultiplet.}
\end{figure}

The brane configurations in Figures 2(a) and 3(a) correspond to $U(1)$ gauge theories with 
one and no charged
hypermultiplets, respectively.  It was noted in Ref.~\cite{HW} that the position
of the D5-brane in the $x^6$ direction does not appear as a parameter in the
gauge theory.  Therefore one could think that there is a phase transition in the
gauge theory when D5 and NS5-branes cross.  In fact, as explained in
Ref.~\cite{HW}, there is no phase transition because a D3-brane is created when
D5 crosses NS5. In the case of the configuration in Figure~2(a), 
moving D5-brane to the right creates a configuration in Figure~2(b).  
The latter corresponds
to a $U(1)$ gauge theory with one hypermultiplet, as there are fundamental strings
connecting the external and internal D3-branes.

\begin{figure}
\setlength{\unitlength}{0.6em}
\begin{center}
\begin{picture}(51,20)
\put(8.5,0.5){\makebox(0,0){(a)}}
\put(4,3){\line(0,1){15}}
\put(11,3){\line(0,1){15}}
\put(4,9){\line(1,0){7}}
\put(13,15){\makebox(0,0){\boldmath $\times$}}

\put(24.5,0.5){\makebox(0,0){(b)}}
\put(21,3){\line(0,1){15}}
\put(28,3){\line(0,1){15}}
\put(21,9){\line(1,0){7}}
\put(24,15){\makebox(0,0){\boldmath $\times$}}
\put(28,15){\line(-1,0){4}}

\put(44.8,0.5){\makebox(0,0){(c)}}
\put(41,3){\line(0,1){15}}
\put(48,3){\line(0,1){15}}
\put(41,9){\line(1,0){7}}

\put(36,15){\makebox(0,0){\boldmath $\times$}}

\put(48,14.8){\line(-1,0){11.8}}
\put(41,15.1){\line(-1,0){4.8}}
\end{picture}\end{center}
\caption{Starting with configuration (a) and moving the D5-brane 
to the left one gets configurations (b) and (c).}
\end{figure}
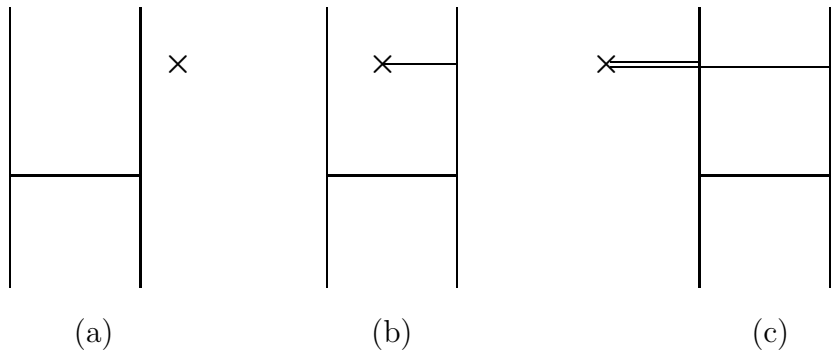

The picture with singular monopoles described in Section~\ref{monsing} refers to
Figure~2(b) on page \pageref{pic}.  Figure~2(a), however, presents a puzzle:  there are no D3-branes
connecting D5 with NS5, so the reasoning of Section~\ref{monsing} seems to imply
that the configuration corresponds to a $U(2)$ monopole on the $A_0$ ALF
space (i.e. the Taub-NUT space) with $n=1,\ell=\ell'=0$. It was explained in the previous
section that
such a monopole is equivalent to a nonsingular monopole on ${\bf R}^3$,
whose moduli space is ${\bf R}^3\times {\bf S}^1$.  This is clearly false, since
the moduli space of a $U(1)$ gauge theory with one charged hypermultiplet is
known to be the Taub-NUT space~\cite{S,SW}, not ${\bf R}^3\times {\bf S}^1$. The resolution
is that Figure~2(a)
corresponds to a monopole with $\ell=1,\ell'=0$, just as in Figure~2(b),
but in a singular gauge.  As explained
in the Appendix, there is a singular $\theta$-dependent gauge transformation
which eliminates the singularity at $\vec{x}=\vec{p}$ but reintroduces one at
the monopole core (the singularity is reflected in the ``hedgehog'' behavior of
the Higgs field near the core). In this new gauge the fields are nonsingular at
$\vec{x}=\vec{p}$, as expected in a situation like Figure 2(a), with no semi-infinite D3-branes
ending on the NS5-branes. Such a ``shaggy monopole'' has the same moduli
space as the normal monopole on the Taub-NUT with $\ell=1,\ell'=0$, which in turn is
equivalent to a monopole on ${\bf R}^3$ with Higgs field diverging near the
point $\vec{x}=\vec{p}$ as in Eq.~(\ref{sing}).  The moduli space of the latter
is indeed the Taub-NUT space~\cite{KrTh}.

What happens if one starts with Figure~3(a) and moves the D5-brane inside?
According to Ref.~\cite{HW} the final configuration must be that in Figure~3(b).
One might expect a charged hypermultiplet from strings connecting the internal
D3-brane with the newly created one.  This again would imply a phase transition
when the D5-brane crosses the NS5-brane.  Moreover, if we move the D5-brane
farther to the left, there will be another D3-brane created (see Figure~3(c)) and
one might think that two hypermultiplets appear!  Is there a phase transition in
this case?

The interpretation in terms of monopoles on Taub-NUT space helps us to
understand what happens to the moduli spaces and to see that there is no phase
transition.  Namely, Figure~3(b) corresponds to a ``shaggy monopole'' with an
additional $\ell=-1,\ell'=0$ singularity at the center of the Taub-NUT space $\vec{x}=\vec{p}$.
The singularities at the monopole core and at $\vec{x}=\vec{p}$ can both be
simultaneously
eliminated by a gauge transformation (see Appendix), and we are back
to the normal monopole with $\ell=0,\ell'=0$.  The latter is equivalent to a nonsingular
$n=1$ monopole on ${\bf R}^3$, and therefore the moduli space is ${\bf
R}^3\times {\bf S}^1$, the same as that of the configuration in Figure~3(a).  In Figure~3(c) 
the D3-branes
connecting the D5 with NS5's correspond to a Dirac monopole embedded in the subgroup $U(1)_{cm}\subset U(2)=U(1)_{cm}\times SU(2)$, therefore they do not influence the $SU(2)$ monopole at
all. (The component of the Higgs field in $U(1)_{cm}$ corresponds to the center-of-mass motion of the fivebranes.)
Thus the moduli space is still ${\bf R}^3\times {\bf S}^1$.

Apparently, the naive counting of string modes fails in
situations like those in Figures 3(b) and 3(c): in these cases there are no stable fundamental string states
connecting the internal D3-branes with the D3-branes stretched between the D5
and NS5-branes.  

Similar arguments apply when there is more than one D3 and/or D5-brane.

\section{Moduli Spaces of Singular Monopoles}\label{moduli}
 
There are several
approaches to finding metrics on monopole moduli spaces.  The most direct one is
to use the Nahm transform~\cite{N}.  In principle, this should yield an isometry
between the monopole moduli space and the space of solutions of Nahm equations.
So far the details have been worked out only for nonsingular $SU(2)$
monopoles~\cite{Nakajima}.  The idea of the minitwistor approach~\cite{H1,H2} is
to encode the monopole data in terms of an algebraic curve in $T{\bf
P}^1$.  This curve is then reinterpreted as a spectral curve of Nahm equations,
in the spirit of Refs.~\cite{Integr,H2}.  This program has been realized for
nonsingular monopoles of all classical groups in Ref.~\cite{HM}.
This approach only allows to prove that
the moduli spaces of monopoles and Nahm data are diffeomorphic.  There is a
natural hyperk\"ahler metric on the space of Nahm data, so it is very plausible
that these manifolds are in fact isometric. 
Here we adopt a less rigorous approach, regarding singular $U(2)$ monopoles as a limit
of nonsingular $SU(3)$ monopoles.  Therefore we can construct the moduli spaces
in question by considering a certain limit of Nahm equations for $SU(3)$
monopoles.

Let us recall what Nahm equations for $SU(3)$ monopoles look like according to
Ref.~\cite{HM}.  In the case of maximal breaking, which is all we need,
$SU(3)$ monopoles are labeled by a pair of nonnegative integers $(n,k)$.  For $n<k$ Nahm
data consist of two quadruplets
\begin{equation}
\left(T_0^{(\lambda)}(s),T_1^{(\lambda)}(s),T_2^{(\lambda)}(s),
T_3^{(\lambda)}(s)\right), \ \lambda=1,2,
\end{equation}
with the
first quadruplet defined for $s\in (0,1)$, and the second one defined for $s\in
(1,\mu)$.  $T_i^{(1)}$ and $T_i^{(2)},\ i=0,\ldots,3$ take values in $u(n)$ and $u(k)$,
respectively.  It is very convenient to combine the functions
$T_i^{(\lambda)},i=0,\ldots,3,\ \lambda=1,2,$ into two quaternions 
\begin{equation}
T^{(\lambda)}= 
T_0^{(\lambda)}+e_1 T_1^{(\lambda)}+e_2 T_2^{(\lambda)}+e_3 T_3^{(\lambda)},
\ \lambda=1,2,
\end{equation}
with $e_i$ being the quaternion units.
(In what follows we will denote the real ($T_0$) and imaginary ($T-T_0$)
parts of quaternions by the symbols $\Rea$ and $\Ima$ respectively, and
think of the purely imaginary quaternions as three-component vectors.)  Thus one
can think of $T^{(1)}$ and $T^{(2)}$ as two functions $T^{(1)}(s):(0,1)\ra
u(n)\otimes{\bf H}$ and $T^{(2)}(s):(1,\mu)\ra u(k)\otimes {\bf H}$.
They must satisfy a number of constraints~\cite{HM}:

(i) Both functions satisfy Nahm equations

\begin{equation}
\frac{d T_i}{ds}+\left[ T_0, T_i\right] = \frac{1}{2}
\vareps^{ijk} \left[ T_j, T_k\right],\; i=1,2,3.  
\end{equation}

(ii) $\Rea\ T^{(1)}(s)$ and $\Rea\ T^{(2)}(s)$ extend smoothly to $[0,1]$ and
$[1,\mu]$, respectively.  $\Ima\ T^{(1)}(s)$ has a simple pole at $s=0$.  The
residue is an $n$-dimensional irreducible representation of $su(2)$.  $\Ima\
T^{(2)}(s)$ has a simple pole at $s=\mu$ with a residue which is a
$k$-dimensional irreducible representation of $su(2)$.

(iii) $\Ima\ T^{(1)}(s)$ extends smoothly to $(0,1]$.  In the neighborhood of
$s=1$ $\Ima\ T^{(2)}(s)$ has the following form

\begin{equation}
\renewcommand{\arraystretch}{1.4} T^{(2)}_i(s)= \left( \begin{array}{cc} \rho_i/s+O(1) &
O\left(s^{(k-n-1)/2}\right) \\ O\left(s^{(k-n-1)/2}\right) & 
T^{(1)}_i(1)+O(s) \end{array}\right), \quad i=1,2,3.
\end{equation}
Here $\rho_i=\rho(i\sigma_i/2),i=1,2,3$, where $\rho$ is a $(k-n)\times(k-n)$
irreducible representation of $su(2)$ and $\sigma_i$ are Pauli matrices.

The set of all Nahm data satisfying conditions (i)-(iii) is invariant with
respect to gauge transformations which act in a more-or-less obvious manner:
the gauge group is $U(n)$ on the interval $[0,1]$ and $U(k)$ on the interval
$[0,\mu]$.  To preserve the condition (iii) and the residues of
$T^{(1)},T^{(2)}$ one must also require that at $s=0$ and $s=\mu$ the gauge
transformations reduce to identity, and at $s=1$ the ``right'' gauge group $U(k)$
reduces to the ``left'' $U(n)$.

For $n>k$ the constraints are the same, with the roles of $T^{(1)}$ and
$T^{(2)}$ interchanged.  For $n=k$ the Nahm data include, in addition, a
quaternionic vector $a\in {\bf H}^n$, and the condition (iii) is replaced by
the following one:

(iii$'$) $\Ima\ T^{(1)}$ and $\Ima\ T^{(2)}$ extend smoothly to $(0,1]$ and
$[1,\mu)$, respectively, so that 
\begin{equation} \Ima\ T^{(2)}_{AB}(1)-\Ima\
T^{(1)}_{AB}(1)=\frac{i}{2} a_{(A} e_1{\bar a}_{B)}+\frac{1}{2} a_{[A} {\bar
a}_{B]}, \ A,B=1,\ldots,n.  
\end{equation} 
(The parentheses and brackets denote symmetrization and
antisymmetrization, respectively.)

In this case the gauge group $U(n)$ acts also on $a$ from the right,
$a_A\ra a_B\ g(1)_{BA}$, where $g(s)$ is a gauge transformation.

The space of all Nahm data modulo gauge transformations is diffeomorphic to the
space of all $(n,k)$ monopoles~\cite{HM}.  There is a natural hyperk\"ahler metric
on the space of Nahm data, and therefore it is expected that the two spaces are
isometric.

To see the metric on the equivalence classes of Nahm data, notice
that the gauge group acts triholomorphically on the flat infinite-dimensional
hyperk\"ahler manifold consisting of {\it all} pairs
$\left(T^{(1)},T^{(2)}\right)$ satisfying (ii) and (iii), except that now the lower
right corner of $\Ima\ T^{(2)}(1)$ need not be equal to $\Ima\ T^{(1)}(1)$.  (For
$n=k$ one must consider instead the space of all triplets
$\left(T^{(1)},T^{(2)},a\right)$ such that $T^{(1)}$ and $T^{(2)}$ satisfy (ii) and
extend smoothly to $s=1$, and $a\in {\bf H}^n$.)  The Nahm equations can be
interpreted as moment map equations for gauge transformations which are identity
at $s=1$.  The boundary conditions for Nahm data at $s=1$ can be interpreted as
moment map equations for the action of the residual gauge group, which is the
group of all gauge transformations modulo those which are the identity at $s=1$.
(This group is $U(min(n,k))$.)  Thus one can use the hyperk\"ahler quotient
construction of Hitchin et al.~\cite{HKLR} to construct a hyperk\"ahler metric on the
space of Nahm data modulo gauge transformations.

To obtain $U(2)$ $n$-monopoles with $k$ singularities of the type $\ell=1,\ell'=0$ one should 
take the limit $\mu\ra\infty$ of $(n,k)$ $SU(3)$ monopoles, fixing the positions of $k$ of them
which become infinitely heavy. In this limit $SU(3)$ is broken down to $U(2)$ at a
very high scale, so that $(0,1)$ monopoles become point-like Dirac monopoles, while $(1,0)$
monopoles remain smooth. It is easy to see that the magnetic charges carried by Dirac monopoles
also come out right. The corresponding brane configuration is shown in Figure~4. 
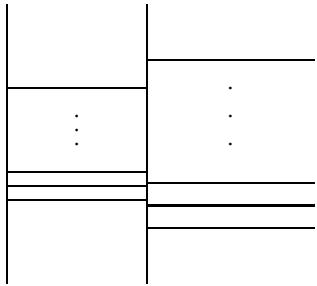
\begin{figure}
\setlength{\unitlength}{0.9em}
\begin{center}
\begin{picture}(17,13)
\put(3,2){\line(0,1){10}}
\put(8,2){\line(0,1){10}}
\multiput(3,5)(0,0.5){3}{\line(1,0){5}}
\put(3,9){\line(1,0){5}}
\multiput(5.5,7)(0,0.5){3}{\circle*{.1}}

\multiput(8,4)(0,0.8){3}{\line(1,0){6}}
\put(8,10){\line(1,0){6}}
\multiput(11,7)(0,1){3}{\circle*{.1}}

\end{picture}
\end{center}
\caption{The brane configuration corresponding to singular $U(2)$
monopoles can be obtained as a limit of that corresponding to 
regular $SU(3)$ monopoles.}
\end{figure}
{}From the above description it is clear that in
this limit $T^{(2)}$ becomes a function defined on $(1,+\infty)$ and satisfying
(i) and (iii) (or (iii$'$) if $n=k$). A natural boundary condition at
$+\infty$ is to require that $\lim_{s\ra +\infty} T^{(2)}(s)$ exist in some gauge,
and that $\Rea\ T^{(2)}(+\infty)=0$. Then Nahm equations imply that the matrices
$T^{(2)}_i(+\infty),\ i=1,2,3,$ commute and can be reduced to a diagonal form $\Ima\
T^{(2)}(+\infty)={\rm diag}(\vec{p}_1,\ldots,\vec{p}_k)$ for some $\vec{p}_\alpha\in
{\bf R}^3$.  It remains to understand what the positions of infinitely heavy
monopoles are.  The Nahm data just described depend on $k$ vectors 
$\vec{p}_\alpha,\ \alpha=1,\ldots,k$.  It is therefore tempting to identify them as the
positions of $k$ infinitely heavy monopoles, i.e.  singularities of the Higgs
field.  This identification can be justified by recalling the physical meaning
of Nahm data~\cite{Diacon}.  The variable $s$ is interpreted as the coordinate
$x^6$ along the horizontal direction in Figure~4.  The matrices $T^{(2)}_i$ describe the transverse coordinates
of semi-infinite D3-branes.  For generic values of $s$ the three matrices
$T^{(2)}_i$ do not commute, so the notion of the transverse position of a given D3-brane is not defined.  However, at $s=+\infty$ the matrices do commute, and
their eigenvalues $\vec{p}_\alpha$ have the meaning of the D3-branes' asymptotic
coordinates in the $(x^3,x^4,x^5)$ plane.  From Section~\ref{alfmon} we know
that these asymptotic coordinates are precisely the positions of the
singularities of the Higgs field.

\section{Examples}\label{example}

In this section we illustrate the above construction by the examples of $U(1)$
and $SU(2)$ gauge theories with $k$ massive fundamentals. In what follows 
$\vec{p}_\alpha,\ \alpha=1,\ldots,k$ will be hypermultiplet masses.
Let us also recall that $n=1$ (one monopole) corresponds to a $U(1)$ gauge theory,
and $n=2$ (two monopoles) corresponds to a $U(2)$ or $SU(2)$ gauge theory.

\subsection{$n=1,\ k$ arbitrary}\label{ex1}

The Nahm data consist of an ${\bf H}\,$-valued function
$T^{(1)}(s)$ on $[0,1]$ and a $u(k)\otimes{\bf H}\,$-valued function
$T^{(2)}(s)$ on $(1,+\infty)$. $T^{(2)}$ has a simple pole at $s=1$, and
the matching condition (iii) is satisfied.
The boundary conditions at $s=+\infty$ are $\Rea\ T^{(2)}(+\infty)=0,\ \Ima\
T^{(2)}(+\infty)={\rm diag}(\vec{p}_1,\ldots,\vec{p}_k)$. 

In this case we expect, from field theory, that the moduli space is the $A_{k-1}$
ALF space~\cite{S}. To obtain this result in our setup we would have
to solve $k\times k$ Nahm equations on a half-line. Alas, we do not know how to do it
directly. Fortunately, there is a way to find the moduli space of Nahm equations
without actually solving them~\cite{Donaldson}. First one splits the three Nahm equations
into one complex and one real equation. This amounts to picking a complex structure
out of a ${\bf P}^1$ of available complex structures. The complex equation is
invariant under the complexification of the gauge group $G^{\bf C}$. Donaldson proved
that the space of solutions of the complex equation modulo $G^{\bf C}$ is the same
as the space of solutions of all three equations modulo $G$. Thus it suffices
to solve the complex equation to determine the moduli space of Nahm equations
as a complex manifold. (This is similar to how one computes the moduli
space of supersymmetric gauge theories: instead of solving both $D$ and $F$-flatness
conditions, one solves only the $F$-flatness conditions modulo the complexification
of the gauge group.) Solving the complex equation is easy since it is locally trivial.
We refer the reader to  Ref.~\cite{Donaldson} for details, and to 
Refs.~\cite{Krcoadjoint,Dancer1,Dancer3} for some applications of this technique.
In our case the moduli space turns out to be isomorphic to a hypersurface
in ${\bf C}^3$ specified by the equation 
\begin{equation}\label{ale}
xy=\prod_{\alpha=1}^{k} (z-p_\alpha),  
\end{equation}
where $p_\alpha$ is the ``complex'' part of $\vec{p}_\alpha$, $p_\alpha=
p_{\alpha 1}+i p_{\alpha 2}$. This is the complex structure of the 
resolution of $A_{k-1}$ singularity.

To find the metric one needs to know all three complex structures, however.
To this end one has to vary the arbitrary complex structure we picked in
the beginning and see how variables in Eq.~(\ref{ale}) change~\cite{Dancer3}. The result is rather simple
to describe: $p_\alpha$ are fixed real sections of ${\cal O}(2)$,
$z$ is the coordinate in the fiber of the line bundle $T{\bf P}^1={\cal O}(2)$
over the
${\bf P}^1$ of complex structures, while $x$ and $y$ take values in the line bundles 
$L(k)$ and $L^{-1}(k)$ over $T{\bf P}^1$, respectively.
(In the notation of Ref.~\cite{H2} $L^x(k)$ is a line bundle over $T{\bf P}^1$ with a
transition function $\zeta^{-k} e^{x \eta/\zeta}$, where $\zeta$ is the
coordinate on ${\bf P}^1$ and $\eta$ is the coordinate in the tangent space at $\zeta$, $d\zeta(\eta)=1$.) 
With such identification Eq.~(\ref{ale}) describes the twistor space
of the $A_{k-1}$ ALF space~\cite{H3}.

\subsection{$n=2,\ k=1$}

The Nahm data consist of a $u(2)\otimes{\bf H}\,$-valued function
$T^{(1)}(s)$ on $(0,1]$ and a $u(1)\otimes{\bf H}\,$-valued function
$T^{(2)}(s)$ on $[1,+\infty)$. $T^{(1)}$ has a simple pole at $s=0$
with residue $e_1\rho_1+e_2\rho_2+e_3\rho_3$, where $\rho_i=i\sigma_i/2,\ i=1,2,3$. The
boundary conditions at $s=+\infty$ are $\Rea\ T^{(2)}(+\infty)=0,\ \Ima\
T^{(2)}(+\infty)=\vec{p}$. At $s=1$ the matching condition (iii) is satisfied.

It proves convenient to perform the hyperk\"ahler quotient in two steps. 
First we take the quotient with respect to gauge transformations which are
the identity at $s=1$. This amounts to solving Nahm equations on two intervals
separately and finding their moduli spaces of solutions. For $s\in (1,+\infty)$ 
Nahm equations just tell us that $\Ima\ T^{(2)}$ is independent of $s$ and equal
to $\vec{p}$, i.e. the moduli space is just a point. Solving Nahm
equations for $2\times 2$ matrices on $(0,1)$ is also elementary,
since the equations can be reduced to those of the Euler top. 
In fact, for $s\in (0,1]$ \ the moduli space of solutions with boundary behavior
as described above has been investigated by Dancer~\cite{Dancer1}.
It turns out that the
moduli space is a 12-dimensional hyperk\"ahler manifold $M^{12}$ of the form
${\bf R}^3\times {\bf S}^1\times M^8$, where $M^8$ is also hyperk\"ahler and
irreducible.  $M^8$ admits a triholomorphic action of $SU(2)$.

Second, we take the quotient with respect to the $U(1)$ group ``living''
at $s=1$. This $U(1)$ is a subgroup of $U(2)$ which is the group of all
gauge transformations modulo those which reduce to identity at $s=1$.
More concretely, the $U(1)$ acts on ${\bf R}^3\times 
{\bf S}^1\times M^8$ as follows: it rotates the ${\bf S}^1$,
and it acts on $M^8$ as a maximal torus of the triholomorphic $SU(2)$ 
mentioned in the end of the previous paragraph. The boundary condition
(iii) implies that the level of the quotient is $2\vec{p}$. The net result
is an 8-dimensional hyperk\"ahler manifold depending on $\vec{p}$ as a parameter.
It is the moduli space of two monopoles with a fixed singularity at 
$\vec{x}=\vec{p}$ and corresponds to the $U(2)$ gauge theory with one massive
fundamental hypermultiplet. 

If one wishes to obtain the Coulomb branch of the $SU(2)$
theory with the same matter content, one should perform a further $U(1)$
hyperk\"ahler quotient (i.e.  pass to the centered monopole moduli space).
This $U(1)$ is easily identified: it acts on ${\bf R}^3\times {\bf S}^1$
by rotating the ${\bf S}^1$. The level of the quotient is simply the position
of the monopoles' center of mass. It can always be absorbed into $\vec{p}$,
so we can set it to zero. Performing this $U(1)$ quotient rids the $M^{12}$ of
the ${\bf R}^3\times {\bf S}^1$ factor. Thus we conclude that the moduli space of the
$SU(2)$ gauge theory with one fundamental is the hyperk\"ahler quotient
of $M^8$ by $U(1)$ at level $2\vec{p}$. This is exactly the four-dimensional manifold
constructed by Dancer in Ref.~\cite{Dancer1} and proposed in Ref.~\cite{SW}
as a candidate for the Coulomb branch. Moreover, we showed above that $\vec{p}$
should be identified as the mass of the hypermultiplet. This agrees with
Ref.~\cite{SW} where it was suggested that the level of the quotient should
be $twice$ the mass of the fundamental.

\subsection{$n=2,\ k=2.$}

In this case the Nahm data consist of a $u(2)\otimes{\bf H}\,$-valued function
$T^{(1)}(s)$ on $(0,1]$, a $u(2)\otimes{\bf H}\,$-valued function
$T^{(2)}(s)$ on $[1,+\infty)$, and a quaternionic vector $a\in {\bf H}^2$.
Both functions satisfy Nahm equations.  $T^{(1)}$ has a simple pole at $s=0$
with residue $e_1\rho_1+e_2\rho_2+e_3\rho_3$, where $\rho_i=i\sigma_i/2,\ i=1,2,3$.  The
boundary conditions at $s=+\infty$ are $\Rea\ T^{(2)}(+\infty)=0,\ \Ima\
T^{(2)}(+\infty)={\rm diag}(\vec{p}_1,\vec{p}_2)$.  At $s=1$ the matching condition
(iii$'$) is satisfied.

Again we split the calculation in two steps. The solution of Nahm equations
for $s\in (0,1)$ is the same as before. To solve the
equations on $(1,+\infty)$, we split $T^{(2)}$ into a part proportional to the
identity matrix and a traceless matrix.  The equations for the ``identity'' part
simply say that $Tr\left(\Ima\ T^{(2)}\right)$ is independent of $s$ and equal to
$\vec{p}_1+\vec{p}_2$.  The equations for the traceless part can be solved in
terms of hyperbolic functions.  After one performs the quotient with respect to
the $U(2)$ gauge group which degenerates to the identity at $s=1$ and to $U(1)\times
U(1)$ at $s=+\infty$, one gets a four-dimensional moduli space $M_{EH}$.  Its
metric can be computed to be the two-center Gibbons-Hawking (or Eguchi-Hanson)
metric with $|\vec{p}_1-\vec{p}_2|$ being the distance between the centers.
Actually, this is a particular case of Kronheimer's construction of hyperk\"ahler
metrics on the coadjoint orbits of a complex group $G$~\cite{Krcoadjoint}.
Kronheimer's construction also uses Nahm equations, and for $G=SL(2,{\bf
C})$ coincides with ours.  (The coadjoint orbit here happens to be
isomorphic to $T{\bf P}^1$ as a complex manifold.)  
The Eguchi-Hanson metric admits a
triholomorphic action of $SU(2)$.

The second step is to take the hyperk\"ahler quotient of 
\begin{equation}
M^{12}\times {\bf H}^2\times M_{EH}
\end{equation}
with respect to $U(2)$.  This residual $U(2)$ is the
quotient of all gauge transformations by those which reduce to the identity at
$s=1$.  The subgroup $U(1)\subset U(2)=U(1)\times SU(2)$ acts on $M^{12}={\bf R}^3\times
{\bf S}^1\times M^8$ by rotating ${\bf S}^1$, and on $a\in {\bf H}^2$ by right
multiplication by $\exp(e_1\phi)$.  The matching condition (iii$'$) means that the
$U(1)$ quotient should be performed at level $\vec{p}_1+\vec{p}_2$.  The $SU(2)$
subgroup acts on the $M^8$ part of $M^{12}$ and on the Eguchi-Hanson space $M_{EH}$.  It also
acts on $a$ by the right multiplication $a\ra a\ g^t$.

The quotient manifold is an 8-dimensional hyperk\"ahler manifold depending on
$\vec{p}_1$ and $\vec{p}_2$ as parameters.  It is the moduli space of two
monopoles with two fixed singularities at $\vec{p}_1$ and $\vec{p}_2$ and
corresponds to the $U(2)$ gauge theory with two massive fundamental
hypermultiplets. 

To obtain the Coulomb branch of the $SU(2)$
theory we must ``center'' the monopoles, as in the previous example. The
position of the center of mass can be set to zero without loss of generality.
As earlier, ``centering'' monopoles is achieved by taking a $U(1)$ quotient.
This procedure eliminates the ${\bf R}^3\times{\bf S}^1$ factor of $M^{12}$.
Then we need to compute the hyperk\"ahler quotient of
\begin{equation}
M^8\times {\bf H}^2\times M_{EH}
\end{equation}
by $U(2)=U(1)\times SU(2)$, where $U(1)$ acts only on ${\bf H}^2$
by right multiplication, and $SU(2)$ acts on all three factors.  The level of the
quotient is $\vec{p}_1+\vec{p}_2$. Since $U(1)$ acts so simply, we can perform
the quotient with respect to it explicitly (see e.g. Ref~\cite{GR}). The final result
is that the the Coulomb branch of the $SU(2)$ theory with two fundamentals is the
hyperk\"ahler quotient of $M^8\times M_{EH}'\times M_{EH}$ with respect to $SU(2)$.
Both $M_{EH}'$ and $M_{EH}$ are the two-center Gibbons-Hawking (Eguchi-Hanson) spaces
with distances between the centers $\vec{p}_1+\vec{p}_2$ and $\vec{p}_1-\vec{p}_2$ respectively. 

\subsection{$n=2,\ k=3$}
 
It is convenient to slightly change our point of view and regard two $U(2)$ monopoles
with three singularities as a limit of $SU(4)$ $(1,2,2)$ monopoles, rather than
the limit of $SU(3)$ $(2,3)$ monopoles.  The limit is such that $(1,\ ,\ )$ and
$(\ ,\ ,2)$ monopoles become infinitely heavy.  The corresponding brane
construction is shown in Figure~5.  
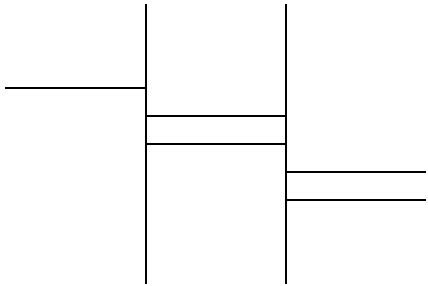
\begin{figure}
\setlength{\unitlength}{0.9em}
\begin{center}
\begin{picture}(17,13)
\put(6,2){\line(0,1){10}}
\put(11,2){\line(0,1){10}}
\multiput(6,7)(0,1){2}{\line(1,0){5}}
\put(6,9){\line(-1,0){5}}
\put(11,6){\line(1,0){5}}

\put(11,5){\line(1,0){5}}

\end{picture}\end{center}
\caption{A $U(2)$ 2-monopole with three singularities
is a limit of a regular $(1,2,2)$ $SU(4)$ monopole.}
\end{figure}
The Nahm data consist of three
functions  $T^{(0)}:(-\infty,0]\ra u(1)\otimes {\bf H},
\ T^{(1)}:[0,1]\ra u(2)\otimes {\bf H},\ T^{(2)}:[1,+\infty)\ra u(2)\otimes {\bf H}$
and a quaternionic vector $a\in {\bf H}^2$.  The boundary conditions at
$s=\pm\infty$ are $\Rea\ T^{(0)}(-\infty)=\Rea\ T^{(2)}(+\infty)=0,\ \Ima\
T^{(0)}(-\infty)=\vec{p}_1,\ \Ima\ T^{(2)}(+\infty)={\rm diag}(\vec{p}_2, \vec{p}_3)$.
Here $\vec{p}_\alpha$ are the positions of the singularities
of the Higgs field.  The matching condition at $s=0$ is
\begin{equation}
\Ima\ T^{(1)}(0)_{2,2}=\Ima\ T^{(0)}(0).
\end{equation}
The matching condition at $s=1$ is
\begin{equation}
\Ima\ T^{(2)}_{AB}(1)-\Ima\ T^{(1)}_{AB}(1)=\frac{i}{2} a_{(A} e_1{\bar a}_{B)}+
\frac{1}{2} a_{[A} {\bar a}_{B]},\ A,B=1,2.
\end{equation}
The advantage of this point of view is that we already know what the moduli
spaces of solutions of Nahm equations look like for $s\in(-\infty,0)$ and $s\in
(1,+\infty)$: in the first instance it is a point, and in the second instance
it is a Eguchi-Hanson space $M_{EH}$ with distance between the centers $|\vec{p}_2-
\vec{p}_3|$. For $s\in (0,1)$ we now have to analyze the space of
solutions of $2\times 2$ Nahm equations with nonsingular boundary behavior.
Luckily, this has also been done by Dancer~\cite{Dancer2}.  The moduli space is a
16-dimensional hyperk\"ahler manifold $N^{16}$ which has the form ${\bf
R}^3\times {\bf S}^1\times N^{12}$. $N^{12}$ is hyperk\"ahler and irreducible.  It admits
a triholomorphic $SU(2)_L\times SU(2)_R$ action. (We call these two $SU(2)$'s $SU(2)_L$
and $SU(2)_R$ because they originate from the action of the residual gauge group at $s=0$
and $s=1$.)

If we perform the hyperk\"ahler quotient
in two steps, as before, on the first step we get $N^{16}\times{\bf H}^2\times M_{EH}$.
On the second step we take the hyperk\"ahler quotient with respect to $U(1)\times U(2)$.
$U(1)$ acts on the $N^{12}$ part of $N^{16}$ by a maximal torus of $SU(2)_L$, and the
level of this quotient is $2\vec{p}_1$. $U(2)=U(1)\times SU(2)$ acts as follows: its $U(1)$ subgroup
 acts only on ${\bf H}^2$ by right multiplication by $\exp(e_1\phi)$, while
its $SU(2)$ subgroup acts on all three factors, the action on $N^{16}$ being that of $SU(2)_R$.
The resulting manifold is the Coulomb
branch of the $U(2)$ gauge theory with three hypermultiplets. ``Centering'' the
monopole moduli space we get the following description of the Coulomb branch of the
$SU(2)$ theory with three hypermultiplets: it is a hyperk\"ahler quotient of 
\begin{equation}
N^{12}\times M_{EH}'\times M_{EH}
\end{equation}
with respect to $U(1)\times SU(2)$, where $M_{EH}'$ and $M_{EH}$
are Eguchi-Hanson spaces with distances between the centers $|\vec{p}_2+\vec{p}_3|$
and $|\vec{p}_2-\vec{p}_3|$, respectively. Here $U(1)$ acts only on $N^{12}$ by a maximal
torus of $SU(2)_L$, and the level is $2\vec{p}_1$.
$SU(2)$ acts on all three factors, the action on $N^{12}$ being that of $SU(2)_R$.

\subsection{$n=2,\ k=4$}

Similarly to the previous example, we regard two monopoles with four singularities as a 
limit of $SU(4)$ $(2,2,2)$ monopoles.  The limit is such that $(2,\ ,\ )$ and
$(\ ,\ ,2)$ monopoles become infinitely heavy.  The corresponding brane
configuration is shown in Figure~6. 
\begin{figure}
\setlength{\unitlength}{0.9em}
\begin{center}
\begin{picture}(17,13)
\put(6,2){\line(0,1){10}}
\put(11,2){\line(0,1){10}}
\multiput(6,7)(0,1){2}{\line(1,0){5}}
\put(6,9){\line(-1,0){5}}
\put(11,6){\line(1,0){5}}
\put(11,5){\line(1,0){5}}
\put(6,4.5){\line(-1,0){5}}


\end{picture}\end{center}
\caption{A $U(2)$ 2-monopole with four singularities
is a limit of a regular $(2,2,2)$ $SU(4)$ monopole.}
\end{figure}
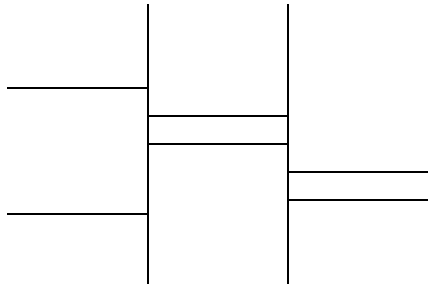

We do not spell out in detail the manipulations
with hyperk\"ahler quotients, since they are almost the same as in the previous example.
We just give the result for the Coulomb branch of the $SU(2)$ gauge theory with four
fundamental hypermultiplets: it is a hyperk\"ahler quotient of 
\begin{equation}
M_{EH}\times M_{EH}'
\times N^{12}\times M_{EH}''\times M_{EH}'''
\end{equation}
with respect to $SU(2)_L\times SU(2)_R$,
where, as the names suggest, $SU(2)_L$ acts on $M_{EH},\ M_{EH}',\ N^{12}$, and
$SU(2)_R$ acts on $N^{12},\ M_{EH}'',\ M_{EH}'''$. The spaces $M_{EH},\ M_{EH}',\ M_{EH}'',$
and $M_{EH}'''$ are Eguchi-Hanson spaces with distances between the centers $|\vec{p}_1+
\vec{p}_2|,\ |\vec{p}_1-\vec{p}_2|,\ |\vec{p}_3+\vec{p}_4|,$ and $|\vec{p}_3-\vec{p}_4|$.   
Arguments from M-theory~\cite{S} or field theory~\cite{SW} show that this space
is an ALF gravitational instanton of type $D_4$. Thus we have a rather simple construction
of such a space.
 
\section{Complex Structures on Moduli Spaces}\label{complex} 

Our description of the metrics on the Coulomb branches is implicit.
This is hardly a drawback, since an explicit formula would be horribly complicated
(see e.g. Ref.~\cite{Dancer1} where the metric corresponding to $n=2,\ k=1$ is discussed).
The only exception is the case of $U(1)$ gauge theories where the moduli space is
of the multi-Taub-NUT form. For general $n$ it is possible to give an explicit description of
the Coulomb branches as {\it complex manifolds.} We remind that hyperk\"ahler manifolds
have three different complex structures $I,\ J,\ K$, but we will concentrate on just
one of them, say $I$. We will limit the discussion of the complex structures
to the cases studied in the previous section.

Computing the complex structures allows us to perform some checks of the metrics.
It follows both from M-theory~\cite{S} and field theory~\cite{SW} considerations
that the Coulomb branch
of $SU(2)$ gauge theory with $k$ fundamentals is an ALF gravitational instanton of
type $D_k$. We will be able to see that indeed our manifolds are resolutions of $D_k$
singularities, at least for $k\leq 4$. A more detailed check can be performed by 
comparison with the Seiberg-Witten solutions of the corresponding $N=2$ $SU(2)$ gauge 
theories
in four dimensions. Recall that in four dimensions the description of the Coulomb branch of
an $N=2$ $SU(2)$ gauge theory involves a complex torus fibered
over a complex plane~\cite{SW4D}. The plane is the moduli space of the theory, while the torus
is an auxiliary object whose $\tau$-parameter is the low-energy gauge coupling.
Upon compactification to three dimensions the total space of this fibration becomes
the moduli space of the corresponding three-dimensional $N=4$ theory~\cite{SW}.
Moreover, it was argued~\cite{SW} that the complex structure remains the same
as in four dimensions (of course, after compactification the moduli space grows
another two complex structures which we disregard in this section.) We will see this
quite explicitly below.

To compute the complex structure on the moduli space of Nahm equations
we followed the approach of Donaldson~\cite{Donaldson} (see Section~\ref{ex1}).
The calculation, although simple in principle, is rather cumbersome and will be
presented elsewhere~\cite{US2}.

\subsection{$SU(2)$ theory with one hypermultiplet}

The complex structure is given by
\begin{equation}\label{13}
y^2=x^2z+2px+1,
\end{equation}
where $p$ is the ``complex'' part of the hypermultiplet mass parameter $\vec{p}$.
The Seiberg-Witten solution is
\begin{equation}\label{14}
y^2=x^2(x-u)+2mx+1,
\end{equation}
where $m$ is the hypermultiplet mass in four dimensions. Obviously, the two complex
structures agree after on sets $z=x-u,\ p=m$ in Eq.~(\ref{13}).

\subsection{$SU(2)$ theory with two hypermultiplets}

The complex structure is given by
\begin{equation}\label{23}
y^2=x^2z-z+2xp_1p_2-(p_1^2+p_2^2).
\end{equation}
The Seiberg-Witten solution is
\begin{equation}\label{24}
y^2=(x^2-1)(x-u)+2xm_1m_2-(m_1^2+m_2^2).
\end{equation}
Eq.~(\ref{23}) agrees with Eq.~(\ref{24}) if one sets $z=x-u,\ p_1=m_1,\ p_2=m_2$.

\subsection{$SU(2)$ theory with three hypermultiplets}

The complex structure is given by
\begin{equation}\label{33}
y^2=x^2z-z^2-z(p_1^2+p_2^2+p_3^2)+2xp_1p_2p_3-(p_1^2p_2^2+p_1^2p_3^2+p_2^2p_3^2).
\end{equation}
The Seiberg-Witten solution is
\begin{eqnarray}\label{34}
y^2&=&x^2(x-u)-(x-u)^2-(x-u)(m_1^2+m_2^2+m_3^2)+2xm_1m_2m_3\nonumber\\
&&-(m_1^2m_2^2+m_2^2m_3^2+m_1^2m_3^2).
\end{eqnarray}
Eqs.~(\ref{33}) and (\ref{34}) agree if one sets $z=x-u,\ p_1=m_1,\ p_2=m_2,\ p_3=m_3$.

\subsection{$SU(2)$ theory with four hypermultiplets}

The complex structure is given by
\begin{eqnarray}\label{43}
y^2&=&x^2z-z^3+z^2\left(S_2-\frac{S_1^2}{2}\right)-z\left(\frac{3S_1^4}{32}-
\frac{3S_1^2S_2}{8}+\frac{S_2^2+S_1S_3}{4}+\frac{S_4}{2}\right)\nonumber\\
&&+x\left(\frac{S_1^4}{32}-\frac{S_1^2S_2}{8}+\frac{S_1S_3}{4}-\frac{S_4}{2}\right)
-\frac{S_1^6}{128}+\frac{3S_1^4S_2}{64}-\frac{S_1^2S_2^2+S_1^3S_3}{16}\nonumber\\
&&+\frac{S_1S_2S_3-S_3^2}{8}+\frac{S_2S_4}{4},
\end{eqnarray}
where $S_\ell$ is an elementary symmetric polynomial in $p_1,\ldots,p_4$
of degree $\ell$, i.e. $S_1=p_1+p_2+p_3+p_4,\ldots, S_4=p_1p_2p_3p_4.$
One immediately sees that this is indeed a resolution of $D_4$ singularity.
To compare with the Seiberg-Witten solution in $D=4$ it is best to
think of $SU(2)$ theory with four flavors as a special case of 
$Sp(2n)$ theory with $2n+2$ flavors. Specializing the formula for the $Sp(2n)$ curve
given in Ref.~\cite{AS}\ to $n=1$ we get the following hyperelliptic curve:
\begin{equation}\label{44}
xy^2=\left(x(x-u)+g\prod_{l=1}^4m_l\right)^2-g^2\prod_{l=1}^4(x-m_l^2).
\end{equation}
Eqs.~(\ref{43}) and (\ref{44}) agree if in Eq.~(\ref{44})
one sets $m_l=ip_l/\sqrt{2g},\ l=1,\ldots,4,$
and makes the following substitution:
\begin{eqnarray}
y&\ra&\frac{y}{\sqrt{g}},\\ \nonumber
x&\ra&\frac{1}{2g}\left(x-z+\frac{S_2}{2}-\frac{S_1^2}{4}\right),\\ \nonumber
u&\ra&\frac{1}{2}\left(x+3z-\frac{S_2}{2}+\frac{S_1^2}{4}\right)+
\frac{1}{2g}\left(x-z+\frac{S_2}{2}-\frac{S_1^2}{4}\right).
\end{eqnarray}

\section{Summary and Conclusions}\label{concl}

In this paper we showed that the Coulomb branches of 
$D=3,\ N=4$\ \ $SU(n)$ and $U(n)$ gauge theories with $k$ fundamental 
hypermultiplets are identical to the (centered) moduli spaces of $n$ $U(2)$ monopoles
with $k$ singularities. We then constructed the latter spaces as infinite
dimensional hyperk\"ahler quotients (i.e. as the moduli spaces
of Nahm equations). This amounts to an implicit description of the exact
metrics on the Coulomb branches. For the simplest cases of $U(1)$ and $SU(2)$ 
gauge theories we also computed the complex structures on the moduli spaces
and compared with expectations from field theory. This provides a check of 
the correspondence between the monopole moduli spaces and the Coulomb branches.
An even more stringent check is afforded by the comparison with Seiberg-Witten
solutions of $D=4,$ $N=2$ theories which upon dimensional reduction yield
our $D=3,$ $N=4$ theories. It was argued in Ref.~\cite{SW} that that a ``distinguished''
complex structure of the $D=3$ moduli space is the same as the complex structure
of the Seiberg-Witten fibration in $D=4$. We checked that this is indeed the case
for $SU(2)$ theory with up to four fundamental hypermultiplets.

As a by-product, we constructed ALF gravitational instantons of type $D_k$ for any positive $k$ 
as centered moduli spaces of two monopoles with $k$ singularities. A detailed discussion
of their metrics and twistor spaces will be presented in forthcoming papers~\cite{us, US2}.
It remains to be seen if these methods can be exploited to construct $E_k$ gravitational instantons.

Finally, using the monopole interpretation, we argued for the absence of
phase transitions in the gauge theory whenever a D5-brane crosses any of the  NS5-branes. It turns
out that brane creation is not sufficient to explain this: in addition
one has to 
postulate that in some brane configurations described in 
Section~\ref{phase}
certain string modes are absent, contrary to naive expectations.

\renewcommand{\thesection}{Acknowledgements}
\section{}

We would like to thank Amihay Hanany, Nigel Hitchin, John H. Schwarz, and Edward Witten for helpful conversations.

\renewcommand{\thesection}{Appendix:}
\section{``Shaggy Monopoles''}

In order to define ``shaggy monopoles'' and give their interpretation in terms
of D-branes, first let us look at self-dual $U(2)$ connections on
${\bf R}^3\times {\bf S}^1$. We can think of these connections as living on the
world-volume of two coincident D6-branes wrapped around ${\bf R}^3\times {\bf S}^1$.
Let $R_A$ be the radius of ${\bf S}^1$, and let $\theta$ be the coordinate along it.
We consider the connections with fixed second Chern class
and fixed conjugacy class of holonomy around ${\bf S}^1$ at infinity. 
If the asymptotic eigenvalues of $A_0$ (the $\theta$-component of the gauge field) are $\mu_1$ and $\mu_2$, the holonomy at infinity is conjugate to

\begin{equation}
W=\left(\begin{array}{cc} e^{2\pi iR_A\mu_1} & 0\\ 0 &  e^{2\pi iR_A \mu_2} \end{array}\right).
\end{equation}
Let's assume that $\mu_1 >\mu_2$, for definiteness.
After one T-dualizes along the $\theta$ direction $2\pi\mu_1$ and $2\pi\mu_2$ are interpreted as the 
$x^6$ positions of the D5-branes (we set the string scale $\alpha'$ to 1). Now with two D5-branes at
points $x^6=2\pi\mu_1$ and $x^6=2\pi\mu_2$ there are two ways to stretch a D3-brane between them. After
T-duality in the $x^6$ both types of stretched D3-branes turn into $U(2)$
monopoles on ${\bf R}^3\times {\bf S}^1$, but with different asymptotic eigenvalues of $A_0$. Namely,  
the eigenvalues are $\mu_1,\mu_2$ in one case and $\mu_2+1/R_A,\mu_1$ in the other~\cite{LeeYi}. Both
configurations are in fact t'Hooft-Polyakov monopoles, with $A_0$ playing the role of the Higgs field,
consequently they are $\theta$-independent.

Of course the eigenvalues of $A_0$ can be changed by a $\theta$-dependent gauge transformation.
For example, to change the eigenvalues from $\mu_2+1/R_A,\mu_1$ to $\mu_1,\mu_2$ one has
to use the following gauge transformation:
\begin{equation}
\label{gg} 
g=\exp\left(\frac{i\theta}{2R_A}\left[\frac{A_0-\frac{1}{2}{\rm tr}A_0}{\norm{A_0-
\frac{1}{2}{\rm tr}A_0}}-1\right]\right),
\end{equation} 
where we defined $\norm{\phi}^2=\frac{1}{2}{\rm tr}\phi^2$. Simultaneously this transformation
makes $A_i$ $\theta$-dependent and singular at every point $\vec{q}$ where the traceless
part of $A_0$ vanishes.  
One can think of such a point as the center of the monopole (there is just one such point
for a single smooth t'Hooft-Polyakov monopole). 
The gauge transformation Eq.~(\ref{gg}) creates a ``hedgehog'' at the monopole
center, in the sense that the eigenvalues of the Higgs field approach $1/R_A,0$ as
$\vec{x}\ra\vec{q}$, but the direction of the Higgs field in the $u(2)$ algebra depends on 
the direction of approach.
This singular and $\theta$-dependent $U(2)$ connection on ${\bf R}^3
\times {\bf S}^1$ is what we call a ``shaggy monopole.'' Naturally, the moduli
space of a ``shaggy monopole'' is the same as the moduli space of a regular
monopole from which it was obtained by a gauge transformation.

We would like to stress
that if only D3-branes of one type are present, one can always use the gauge
in which the corresponding gauge configuration is $\theta$-independent. But
if D3-branes stretched both ways are present, then $\theta$-dependence cannot
be removed by a gauge transformation. In particular the "Wilson-line instanton"
of Ref.~\cite{LeeYi} is $\theta$-dependent.

Now we consider a monopole of magnetic charge 1 on the $A_{k-1}$ ALF space. Let the
asymptotic radius of the compact direction be $R_A$. As in the case of 
${\bf R}^3\times {\bf S}^1$ we want to fix the holonomy at infinity.
Monopoles on the $A_{k-1}$ ALF space can be obtained from smooth 
BPS monopoles on ${\bf R}^3$ with asymptotic eigenvalues
of the Higgs field at infinity either $\mu_1,\mu_2$ or 
$\mu_2+1/R_A,\mu_1$, the holonomy being the same. Let us choose the latter possibility.
Performing the change of variables
Eq.~(\ref{change}) we get a $\theta$-independent connection $\tilde{A}$ on the
$A_{k-1}$ ALF space. Since $V^{-1}\left(\vec{p}_\alpha\right)=0$ for all $\alpha$,
the traceless part of $\tilde{A}_0$ vanishes not only at the monopole center, but also at the $k$
centers of the $A_{k-1}$ ALF space. Now suppose we want to change the asymptotic eigenvalues
of $\tilde{A}_0$ to $\mu_1,\mu_2$. To this end we perform the gauge transformation
as in Eq.~(\ref{gg}). The new connection will be singular at the monopole center, as well
as at $\vec{x}=\vec{p}_{\alpha},\alpha=1,\ldots,k.$ 
Thus a smooth monopole on the $A_{k-1}$ ALF space with all $\ell_\alpha$ equal to zero
is gauge-equivalent to a ``shaggy monopole'' which has
singularities at the monopole core and at the centers of the $A_{k-1}$ ALF space.

On the other hand we can start with a $\theta$-independent monopole on the
$A_{k-1}$ ALF space which has $\ell_\alpha=1,\ell'_\alpha=0,\ \alpha=1,\ldots,k.$. 
After the gauge transformation inverse to that in Eq.~(\ref{gg}) it turns into a connection
which has $\hat{A}_0(\vec{p}_{\alpha})=0,\ \alpha=1,\ldots,k$, and a ``hedgehog''
in the monopole center. (Notice that
in Section~\ref{monsing} we set $R_A=2$.) Thus we can trade the singularities at the centers
of the $A_{k-1}$ ALF space for a similar singularity at the monopole center
by means of a singular gauge transformation.

\end{document}